\documentclass[pra,
                twocolumn,
                preprintnumbers,
                amsmath,
                amssymb,
                showpacs,
                superscriptaddress]{revtex4-1}
\usepackage{graphicx} 
\usepackage{dcolumn}  
\usepackage{bm}       
\usepackage{amsfonts}
\usepackage{dsfont}
\usepackage{mathtools}
\usepackage[colorlinks=true,linkcolor=blue,citecolor=red,urlcolor=blue]{hyperref}
\usepackage{color}
\usepackage{natbib}
\newcommand{\ful}{\mbox{C$_{60}$}}
\newcommand{\eq}[1]{Eq.~(\ref{#1})}
\newcommand{\ri}{\mbox{$R_{\mbox{\scriptsize in}}$}}
\newcommand{\ro}{\mbox{$R_{\mbox{\scriptsize out}}$}}
\newcommand{\vscf}{\mbox{$\delta V$}}

\begin{document}

\title{Fullerene photoemission time delay explores molecular cavity in attoseconds}

\author{Maia Magrakvelidze}
\affiliation{%
Department of Natural Sciences, D.\ L.\ Hubbard Center for Innovation and Entrepreneurship,
Northwest Missouri State University, Maryville, Missouri 64468, USA}

\author{Dylan M. Anstine}
\affiliation{%
Department of Natural Sciences, D.\ L.\ Hubbard Center for Innovation and Entrepreneurship,
Northwest Missouri State University, Maryville, Missouri 64468, USA}

\author{Gopal Dixit}
\affiliation{%
Max Born Institute, Max-Born-Strasse 2A, 12489 Berlin, Germany}

\author{Mohamed El-Amine Madjet}
\affiliation{%
Qatar Environment and Energy Research Institute (QEERI), P.O. Box 5825, 
Doha, Qatar}

\author{Himadri S. Chakraborty}
\affiliation{%
Department of Natural Sciences, D.\ L.\ Hubbard Center for Innovation and Entrepreneurship,
Northwest Missouri State University, Maryville, Missouri 64468, USA}

\date{\today}



\begin{abstract}
Photoelectron spectroscopy earlier probed oscillations in $\ful$ valence emissions, producing
series of minima whose energy separation depends on the molecular cavity. We show here that the quantum 
phase at these cavity minima exhibits variations from strong electron correlations in $\ful$, 
causing rich structures in the emission time delay. Hence, these minima offer unique spectral zones to
directly explore multielectron forces via attosecond RABITT interferometry 
not only in fullerenes, but also in clusters and nanostructures for which such minima are likely abundant.  
\end{abstract}

\maketitle 

\section{Introduction}

Resolving electron dynamics in real-time offers the access into a plethora of electron-correlation driven processes in 
atomic, molecular and more complex systems. Advent in technology in producing isolated ultrashort laser pulses 
and pulse trains dovetails a new landscape of active and precision research of light-matter interactions on ultrafast time scale~\cite{hentschel2001, corkum2007, krausz2009, sansone2012}.  
For instance, in pump-probe laser spectroscopy, a pump pulse initiates an electronic process while a subsequent probe pulse  
explores the electron's motion with
a temporal resolution of a few femtoseconds to several attoseconds. This serves as a microcosm of a fundamental
mechanism that a laser-driven process can be viewed as dynamical electronic wavepackets with evolving amplitudes,
phases and group delays. 

Relative delay between 2$s$ and 2$p$ photoemission in neon was measured in a pilot experiment 
by attosecond streaking metrology~\cite{schultze2010delay}. 
Also, for argon, the relative delay between 3$s$ and 3$p$ photoemissions at energies below
the 3$s$ Cooper minimum~\cite{klunder2011probing,guenot2012photoemission} and the group delay in 3$p$ photorecombination 
across the 3$p$ Cooper minimum~\cite{schoun2014} were accessed using attosecond interferometry, known as RABITT. For simple molecules like diatomic nitrogen, 
two-color photoionization, resolved in attoseconds, was the subject of a recent study~\cite{caillat2011}.
Moving to the other extreme in the structure scale, the condensed-phase systems, recent activities
include measurements of the relative delay between the emission from conduction and valence 
band states of monocrystalline magnesium~\cite{neppl2012attosecond} and tungsten~\cite{cavalieri2007attosecond}. 
Further, theoretical studies to explore delays in photoelectrons from metal surfaces brought about important insights~\cite{zhang2009metal}.

Straddling the line between atoms and condensed matters
are clusters and nanostructures that not only have hybrid properties of the two extremes, but also exhibit special behaviors
with fundamental effects and technological applications. Time-resolved access into the photoemission 
processes in fullerenes can be singularly attractive
due to their eminent symmetry and stability. 
Recent efforts were made to predict the time delay in photoemissions from atoms endohedrally confined in $\ful$~\cite{pazourek, dixit2013, deshmukh2014}.  
However, these studies did not address the direct response of $\ful$ electrons, but instead focused at the effects of
confinement. 
Only recently, an electron momentum imaging measurement is performed to 
study the photoelectron angular distribution of $\ful$, establishing an indirect connection to 
the emission time delay at the plasmon resonance~\cite{barillot2015}.
Evidently, hardly anything has been done to temporally explore cluster systems. 
In this Letter, we report
an investigation of the time delay in photoemission from two highest occupied molecular orbitals, HOMO and \mbox{HOMO-1}, of 
$\ful$ which uncovers dramatic attosecond response
at characteristic emission minima. Results carry signatures of $\ful$ cavity, opening a new approach for molecular imaging applications,
and most importantly establish an attosecond route to probe a remarkable aspect of electron correlations.

\section{Essential details of the method}

Time-dependent local 
density approximation (TDLDA) is employed to simulate the dynamical response of $\ful$ to incident photons~\cite{madjet2008}.
The dipole interaction, $z$, with the light that is
linearly polarized in $z$-direction induces a frequency-dependent complex change in the electron 
density arising from dynamical electron correlations. This can be written, using the independent particle (IP) susceptibility $\chi_0$, as
\begin{equation}\label{ind_den2}
\delta\rho({\bf r};\omega)={\int \chi_0({\bf r},{\bf r}';\omega) 
                           [z' + \vscf({\bf r}';\omega)] d{\bf r}'},
\end{equation}
in which
\begin{equation}\label{v_scf}
\vscf({\bf r};\omega) = \int\frac{\delta\rho({\bf r}';\omega)} {\left|{\bf r}-{\bf r}'\right|}d{\bf r}'
     +\left[\frac{\partial V_{\mbox{xc}}}{\partial \rho}\right]_{\rho=\rho_{0}} \!\!\!\!\delta\rho({\bf r};\omega),
\end{equation}
where the first and second term on the right hand side are, respectively, the induced change of the Coulomb and the 
exchange-correlation potentials. Obviously, $\vscf$ 
includes the dynamical field produced by important electron correlations within the linear response regime.

\begin{figure}[h!]
\includegraphics[width=9cm]{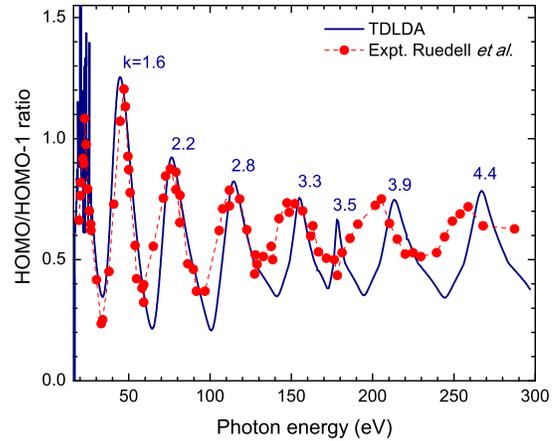}
\caption{(Color online) Oscillations in the ratio of HOMO to \mbox{HOMO-1} photoemission cross sections of $\ful$ calculated using TDLDA/LB94 and compared with the experimental ratio~\cite{ruedel2002oscExp}. 
Similar in~\cite{ruedel2002oscExp}, a smooth background, that roughly fits the total photoionization cross section of atomic carbon~\cite{taylor1976}, is added to TDLDA cross sections to approximately account for local scatterings from carbon atoms. Photoelectron momenta at the peaks are indicated to illustrate 
that the conjugate of the oscillation period relates to the $\ful$ diameter. 
}\label{fig1}
\end{figure}

The gradient-corrected Leeuwen and Baerends exchange-correlation functional [LB94]~\cite{van1994exchange} 
is used for the accurate asymptotic behavior of the ground state potential. 
The $\ful$ molecule is modeled by smearing sixty C$^{4+}$ ions into a spherical jellium shell, fixed in space, with 
an experimentally known $\ful$ mean radius 
($R$ = 3.54 \AA) and a width ($\Delta$ = 1.3 \AA) determined {\em ab\,initio}~\cite{madjet2008}. Inclusion of molecular orientations will have minimal effect on the result due to the 
$\ful$ symmetry~\cite{ciappina2007-2008}.
The delocalized system 
of total 240 valence electrons from sixty carbon atoms constructs the ground state in the Kohn-Sham frame~\cite{madjet2008} using LB94. 
This produced HOMO and \mbox{HOMO-1} to be of 2$h$ ($l=5$) and 2$g$ ($l=4$)
character respectively with each having a radial node -- a result known from the quantum chemical calculation~\cite{troullier1992} supported by direct
and inverse photoemission spectra~\cite{weaver1991}, and from energy-resolved electron-momentum density measurements~\cite{vos1997}. 
TDLDA predicted oscillatory photoemission cross sections 
of  HOMO and \mbox{HOMO-1} in $\ful$ 
which agreed well with the experiment~\cite{ruedel2002oscExp} and with quantum chemical calculations~\cite{ruedel2002oscExp,korica2010}.
Fig.~\ref{fig1} shows a very good agreement between measurements and TDLDA  
ratio of HOMO and \mbox{HOMO-1} cross sections for the four low energy oscillations. 
An extra peak at 175 eV for TDLDA, and a slight offset between the theory-experiment positions of two high-energy peaks, {\em plus} some mismatch between their
widths, are likely limitations of the jellium core. 
These oscillations are due to the interference between emissions from $\ful$ shell-edges 
as was shown by Fourier transforming the above ratio~\cite{ruedel2002oscExp,mccune2008} and evident
from the fact in Fig.~\ref{fig1} that the reciprocal, $2\pi/\Delta k$, of the average peak separation 
($\Delta k \sim 0.5$ a.u.) in photoelectron momentum ($k$) roughly equals the fullerene diameter. The comparison gives confidence on the use of LB94.

Similar geometry-based oscillations in high-harmonic spectra of icosahedral fullerenes were 
predicted~\cite{ciappina2007-2008}. This points to a common spectral implication between photoionization and recombination matrix elements.  

\section{Cavity minima}

Studies of ionization time delay at resonances and minima (anti-resonances) are attractive, since electron correlations can directly influence the result. Of
particular interest is a Cooper minimum which arises at the zero of the wave function overlap in the 
matrix element when the bound wave contains at least one radial node~\cite{cooper1962photoionization}. 
Around this minimum, the ionization probability is diminished which allows couplings with other electrons to dominate, offering a unique spectral zone to probe the correlation.
We show that the minima in the oscillation of $\ful$ valence emissions also appear
from zeros in the matrix element, and thus can be of great value in capturing time-resolved many-electron dynamics.    

\begin{figure}[h!]
\includegraphics[width=9cm]{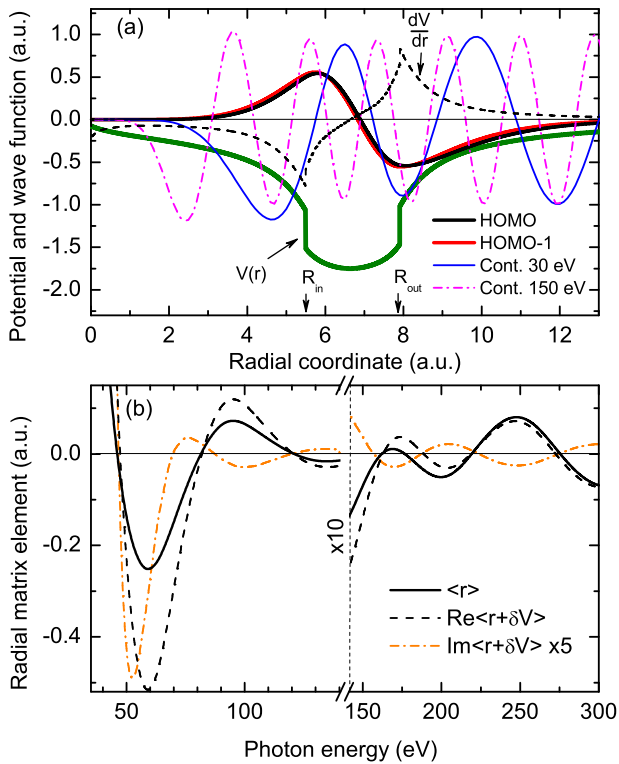}
\caption{(Color online). (a) Ground state radial potential and its gradient, radial wave functions of HOMO and \mbox{HOMO-1}, and the continuum
wave of ($l+1$) angular momentum for a low and a high energy are shown. (b) The real IP radial matrix element 
compares to the real and imaginary components of complex TDLDA matrix element. Besides two scaling regions, the imaginary part is multiplied by an overall factor of 5.} \label{fig2}
\end{figure}

Choosing the photon polarization along $z$-axis, the photoionization dipole amplitude in the IP picture, 
that omits the electron correlation dynamics, is $d = \langle \Psi_{kl'} | z | \Phi_{nl} \rangle$ in which 
$\Phi_{nl}=\phi_{nl}(r)Y_{lm}(\mathbf{\Omega_{r}})$ is HOMO or \mbox{HOMO-1} wave function, and 
the continuum wave function with $l'=l\pm1$ is
\begin{equation}\label{cont}
\Psi_{kl'}({\bf r}) = (8 \pi)^{\frac{3}{2}} \sum_{m'} e^{i \eta_{l'}} \psi_{k l'}(r) 
Y_{l' m'}(\mathbf{\Omega_{r}}) Y_{l' m'}^{*}(\mathbf{\Omega_{k}}),
\end{equation}
where the phase $\eta_{l'}(k)$ includes contributions from the short range and Coulomb potentials, besides a constant $l'\pi/2$.
Using \eq{cont}, the radial matrix element (in length gauge) embedded in $d$ is $\langle r \rangle = \langle \psi_{kl'} | r | \phi_{nl} \rangle$. 
This matrix element can also
be expressed in an equivalent {\em acceleration gauge} as $\langle \psi_{kl'} | dV/dr | \phi_{nl} \rangle$, which embodies the notion
that an ionizing (recoil) force $dV/dr$ is available to an electron in a potential $V(r)$. Both $\ful$ radial ground state potential and its derivative 
are shown in Fig.~\ref{fig2}(a). The potential exhibits rapid variations at the inner ($\ri$) and outer ($\ro$) radii but has
a flatter bottom. Consequently,
the derivative peaks (or anti-peaks) at the shell-edges, allowing two dominant contributions in the 
integral so one can approximate the matrix element as~\cite{mccune2008}
\begin{equation}\label{matelm}
\langle r \rangle \approx A(k) [a_{\mbox{\scriptsize in}} \psi_{kl'}(\ri) + a_{\mbox{\scriptsize out}} \psi_{kl'}(\ro)],
\end{equation}
where $a_{\mbox{\scriptsize in}}$ and $a_{\mbox{\scriptsize out}}$ are the values of $\phi_{nl}$ at $\ri$ and $\ro$, and $A(k)$ is a decaying 
function of $k$ similar to the one calculated semi-classically for metal clusters~\cite{frank1996}. In essence, this
means a strong cancellation effect in the matrix elements at the interior region of the potential due to overlaps between oscillating $\psi_{kl'}$ and radially symmetric $\phi_{nl}$.
This symmetry, not present in atoms (where electrons are localized toward the nucleus), is a character of nanosystems with delocalized electrons; 
see the HOMO and \mbox{HOMO-1} wave functions in Fig.~\ref{fig2}(a). In any case, each term in \eq{matelm}
oscillates in $k$ and vanishes when a node of $\psi_{kl'}$ moves through $\ri$ or $\ro$ or, equivalently, 
when an integer number of half-periods of continuum oscillation fits within $\ri$ or $\ro$; decreasing period of continuum waves with increasing
energy is illustrated in Fig.~2(a). For each term, the effect is analogous to a single spherical-slit diffraction. Since the combination (interference)
of two oscillations is itself an oscillation, $\langle r \rangle$ must also contain zeros, as shown in Fig.~\ref{fig2}(b) for HOMO $\rightarrow k\,(l+1)$.
Evidently, unlike the zero of a Cooper minimum, that depends on the node in the bound wave function, these zeros arise from nodes in the continuum wave function and can be
termed as the {\em cavity minima}. 

\section{Results and discussion}

Wigner-Smith time delay, the energy differential of the
phase of the photoemission amplitude~\cite{wigner1955smith1960}, is accessible by ``two-color" XUV-IR schemes like attosecond streak camera and RABITT. 
This is because
the extra delay introduced by the IR probe pulse, the Coulomb-laser coupling delay,
can be independently calculated and deducted from the data~\cite{pazourek,ivanov2011accurate,dahlstrom2012theory}. Our results~\cite{magrakvelidze2014}
of Wigner-Smith delay using the current TDLDA/LB94 scheme showed excellent agreements with RABITT measurements~\cite{klunder2011probing,guenot2012photoemission,schoun2014} 
for argon. This standard techniques to extract the IR-induced delay information from the Coulomb and the short-range potentials are well described within the IP frame~\cite{ivanov2011accurate}. Ref.~\cite{dahlstrom2012theory} derives this coulomb-laser coupling delay from a universal phase brought by the absorption of the IR photon in the presence of the Coulomb potential with charge Z. Whether multielectron effects like configuration interactions from level compactness could modify this delay is only a question for future research. In fact, experimental efforts to measure the current predictions or those from Ref.~\cite{barillot2015} can only verify the validity of this question.

The IP radial matrix element $\langle r \rangle$ is real, implying that 
the IP phase is directly $\eta$ in \eq{cont} and, hence, insensitive to the zeros in the matrix element. 
However, the phase becomes sensitive to the cavity minima when TDLDA includes correlations via an energy-dependent complex induced potential $\delta V$
in the amplitude: $D = \langle \Psi_{kl'} | z + \delta V({\bf r})| \Phi_{nl} \rangle$; see Ref.~\cite{madjet2008} for details of the formalism.
Hence, the many-body effects could be directly probed by the phase and group delay measurements at these minima. 
The TDLDA phase   
\begin{equation}\label{phase}
\gamma \!=\! \eta \!+\! \arctan\!\! \left[\!\frac{\mbox{Im}\langle r \!+\! \delta V \rangle}{\mbox{Re}\langle r \!+\! \delta V\rangle} \!\right] \!\!=\!
\eta \!+\! \arctan\!\! \left[\!\frac{\mbox{Im}\langle \delta V \rangle}{\langle r \rangle \!+\! \mbox{Re}\langle \delta V \rangle} \!\right],
\end{equation}
since $\langle r \rangle$ is real. In \eq{phase}, the new radial matrix element $\langle r+\delta V(r) \rangle$ 
being complex suffers a $\pi$ phase-shift as its real part moves through a zero at a cavity minimum.

\begin{figure}[h!]
\includegraphics[width=9cm]{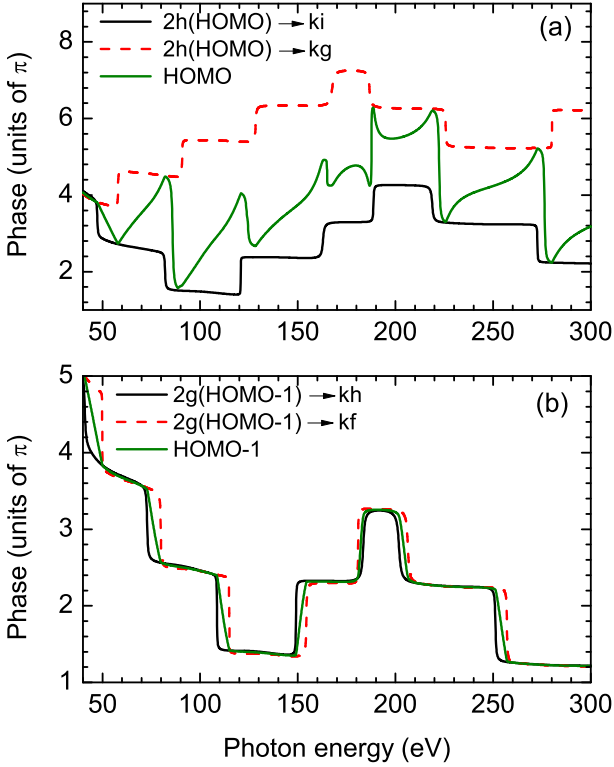}
\caption{(Color online). TDLDA phases as a function of the photon energy for ionization through dipole allowed channels for (a) HOMO and (b) HOMO-1 electrons.
Calculated ionization total phases from these levels are also shown.} \label{fig3}
\end{figure}
\begin{figure}[h!]
\includegraphics[width=9cm]{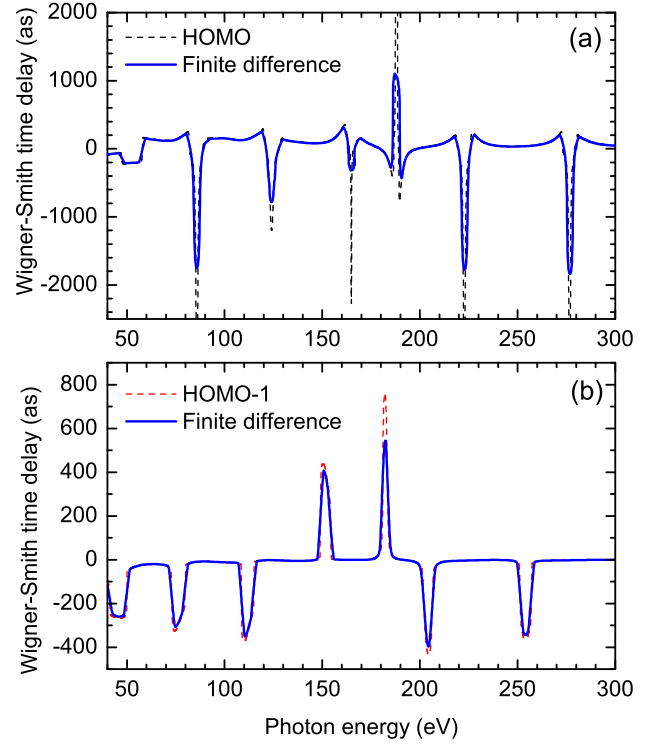}
\caption{(Color online). Wigner-Smith time delays for (a) HOMO and (b) HOMO-1 calculated  within TDLDA framework and its comparison with 
the delays determined by a finite difference approach, where 800 nm IR pulse is used for energy differential (see the text).} \label{fig4}
\end{figure}

TDLDA quantum phases for two dipole channels from each of HOMO and HOMO-1 are 
presented in Figs.~\ref{fig3}. Phase-shifts of about $\pi$ at all cavity minima are noted; 
for HOMO-1 the shifts are roughly synchronized between the two channels [Fig~\ref{fig3}(b)]. 
The direction of a phase-shift, upwards or downwards, depends on the details of the TDLDA matrix element. 
\eq{phase} suggests that $\langle r \rangle$ is 
correlation-corrected by Re$\langle \delta V \rangle$, but this correction diminishes at higher energies as seen in Fig.~\ref{fig2}(b).
When Re$\langle r+\delta V \rangle$ sloshes through a zero, a $\pi$-shift occurs. But the direction
of the shift depends on the sign of Im$\langle r+\delta V \rangle$ -- a quantity {\em entirely} correlation-induced. The oscillations 
in the imaginary part [Fig.~\ref{fig2}(b)] arise from a multichannel coupling with a large number of $\ful$ inner channels which are open at these energies.
The amplitudes of these inner channels do not oscillate in-phase and have diverse phase offsets in relation to HOMO and HOMO-1 oscillations~\cite{mccune2008}.
Consequently, the position of zeros in Im$\langle r+\delta V \rangle$ is a function of correlations via this multichannel process.  
Indeed, while the real and imaginary components are
seen to oscillate roughly out-of-phase, the zeros of one do not occur systematically on a definite side of the zeros of the other, causing
the phase change to follow a pattern 
that directly maps the correlation the valence emission experiences at a cavity minimum.

In the RABITT experiment, one measures the delay associated with a phase $\Gamma$ which is not resolved in the photoemission direction $\Omega_k$:
\begin{equation}\label{total-phase}
\Gamma = \arg [\bar{{\cal D}}_{l+1}\exp(i\gamma_{l+1}) + \bar{{\cal D}}_{l-1}\exp(i \gamma_{l-1})],
\end{equation}
where $\bar{{\cal D}}_{l'} = \int d\mathbf{\Omega_k} |D_{l'}(\mathbf{\Omega_k})|$. Since for a channel $\sigma_{l'} \sim \int d\mathbf{\Omega_k} |D_{l'}(\mathbf{\Omega_k})|^2$, 
we approximate \eq{total-phase} by replacing $\bar{\cal D}$ by the square root of respective channel cross sections.
Fig.~\ref{fig3} also presents these calculated total phases.

TDLDA Wigner-Smith time delays, energy-differentials of total phases, are shown in Fig.~\ref{fig4}.  
To obtain the delay from the phases, one can use arbitrarily small energy steps for the differential. 
Measurements based on RABITT metrology typically uses 800 nm ($\omega$= 1.55 eV) IR probe pulse that leads to the extraction of the delay from measured $\Gamma$ by
$[\Gamma(E+\omega) - \Gamma(E-\omega)]/2\omega$.      
Resulting ``finite difference" TDLDA delays for HOMO and HOMO-1 are also shown in Figs.~\ref{fig4}. 
Structures,
corresponding to negative or positive delays, at the cavity minima indicate striking variations in the photoelectron
speed. The fast (slow) emissions are effects of dynamical anti-screening (screening) from
the multichannel coupling based on the Fano scheme~\cite{fano1961}. In this, the correlation $\langle \delta V({\bf r}) \rangle$
for the emission from $nl$ (HOMO or HOMO-1) reads as~\cite{madjet2008}
\begin{equation}\label{multichannel}
\langle \delta V \rangle _{nl}  =  
 \displaystyle\sum_{\lambda} \lim_{\delta \rightarrow 0}\int dE' \frac{\langle\tilde{\psi}_{\lambda}(E')|\frac{1}{|{\bf r}_{nl}-{\vec{r}}_{\lambda}|}
|\tilde{\psi}_{nl}(E)\rangle}{E-E'+i\delta} d_{\lambda} (E'),
\end{equation}
where the sum is over all other open channels $\lambda$ and two-body wave functions $\tilde{\psi}$ involve both bound and continuum states in an IP channel.
$\langle \delta V \rangle$ can be large, since bound wave functions of delocalized electrons occupy similar regions in space enabling large overlap in \eq{multichannel}.
We note that the details of the correlation here is pretty complex, as all the open channels (about 30 in a jellium frame), constituting 240 delocalized electrons, are coupled.
A simple interpretation of the results may still be outlined. At an XUV energy of current interest, each molecular level can ionize in its uncoupled IP channel. However,
the interchannel coupling in \eq{multichannel} may include another possibility: An inner electron can initially absorb the XUV photon and then transfer the energy via Coulomb interactions 
to HOMO or HOMO-1 to cause an outer emission.
Thus, since this repulsive $1/r_{12}$ underpins
the coupling landscape [\eq{multichannel}], and since the correlation must dominate near a minimum of a channel, either of the valence electrons
feels a strong outward force, via interchannel couplings, from the host of inner electrons and hence ionize faster. This is seen in Fig.~\ref{fig4} in predominant negative-delay structures. The exception at
190 eV needs further investigation. The HOMO-1 level, being below HOMO, feels some blockade from the inward Coulomb push via its coupling with the outer HOMO, and therefore gets relatively slower
overall and, in particular, shows a second positive delay at 150 eV.

Probing correlation forces by the attosecond spectroscopy is the main focus of this work. 
Even though the separation between HOMO and HOMO-1 is 1.3 eV, our results can be experimentally accessed, 
since the resolution of RABITT measurements is not limited to the spectral-width of the attosecond pulse
but to that of the individual harmonics ($\sim$100 meV) of the resulting frequency comb.
Further, by approximating $\psi$ by the asymptotic form $\cos (kr-l'\pi/2)$ of the spherical Bessel function~\cite{landau}, \eq{matelm}
becomes sinusoidal in $k$. This results oscillations in the momentum space with radii being the frequencies. 
Hence, the reciprocals ($\pi/\Delta k$) of the separations (periods) $\Delta k$ between the minima, or between the delay extrema, connect to $\ful$ radii. 
Obviously, for larger (smaller) fullerenes the structures will compactify (spread out). Furthermore, this technique may apply to access time information 
in a spheroidal fullerene, a carbon nanotube, or nanostructures 
of partial symmetry by properly orienting the polarization of XUV photon to minimize non-dipole effects from deformity~\cite{chakraborty}. 

The utilization of plane waves, instead of the continuum solutions as \eq{cont}, should produce cavity minima in the cross section,
since, as discussed above, the origin of these minima is the nodes in the photoelectron wave that plane waves have. But in this case, the minima will appear at spectral positions different from the present result.
Futhermore, as plane waves omit the Coulomb and short range phases of \eq{cont}, the phase and time-delay profiles will differ
from the current prediction. The plane waves routinely form the basis of the strong field approximation. But since the correlation effects
diminish in a strong field environment, the delay structure may considerably weaken or be altered directly by the field.
        
\section{Conclusion}

In summary, photoemission quantum phases and Wigner-Smith time delays for HOMO and HOMO-1 electrons of a $\ful$ molecule 
are investigated. Results show structures at the cavity minima in the energy range above the plasmon resonances and below the carbon $K$-edge
which carry the direct imprint of the dynamical correlation and the molecular size. 
Even though a jellium description of the ion core
omits the scattering from local carbon ions~\cite{madjet2008}, the structures should still be observed, but may soften 
in strength. We also calculated the results with a different, but less accurate than the current (LB94), XC functional. In specific, using a functional as in Ref.~\cite{gunnarsson1976} has shown similar qualitative results. We plan to include the comparison in a future paper. Besides fullerenes, the detection of photoemission minima in metal clusters~\cite{jaenkaelae2011} suggests a possible universality of the phenomenon 
in cluster systems, or even quantum dots~\cite{chakraborty}, that confine finite-sized electron gas. The work predicts a new research direction to apply
attosecond RABITT metrology in the world of gas-phase nanosystems.

\begin{acknowledgments}
The research is supported by the NSF, USA.
\end{acknowledgments}


\begin{thebibliography}{10}

\bibitem{hentschel2001}
M.~Hentschel, R.~Kienberger, C.~Spielmann, G.~A. Reider, N.~Milosevic,
  T.~Brabec, P.~Corkum, U.~Heinzmann, M.~Drescher, and F.~Krausz,
\newblock Nature {\bf 414}, 509 (2001).

\bibitem{corkum2007}
P.~B. Corkum and F.~Krausz,
\newblock Nature Physics {\bf 3}, 381 (2007).

\bibitem{krausz2009}
F.~Krausz and M.~Ivanov,
\newblock Rev. Mod. Phys. {\bf 81}, 163 (2009).

\bibitem{sansone2012}
G.~Sansone, F.~Calegari, and M.~Nisoli,
\newblock Journal of Selected Topics in Quantum Electronics, {\bf 18}, 507 (2012).

\bibitem{schultze2010delay}
M.~Schultze, M.~Fieß, N.~Karpowicz, J.~Gagnon, M.~Korbman, M.~Hofstetter,
  S.~Neppl, A.~L. Cavalieri, Y.~Komninos, T.~Mercouris, C.~A. Nicolaides,
  R.~Pazourek, S.~Nagele, J.~Feist, J.~Burgdörfer, A.~M. Azzeer, R.~Ernstorfer,
  R.~Kienberger, U.~Kleineberg, E.~Goulielmakis, F.~Krausz, and V.~S. Yakovlev,
\newblock Science {\bf 328}, 1658 (2010).

\bibitem{klunder2011probing}
K.~Kl{\"u}nder, J.~M. Dahlstr\"{o}m, M.~Gisselbrecht, T.~Fordell, M.~Swoboda,
  D.~Guénot, P.~Johnsson, J.~Caillat, J.~Mauritsson, A.~Maquet, R.~Taïeb, and
  A.~L'Huillier,
\newblock Physical Review Letters {\bf 106}, 143002 (2011).

\bibitem{guenot2012photoemission}
D.~Gu{\'e}not, K.~Kl{\"u}nder, C.~L. Arnold, D.~Kroon, J.~M. Dahlstr{\"o}m,
  M.~Miranda, T.~Fordell, M.~Gisselbrecht, P.~Johnsson, J.~Mauritsson,
  E.~Lindroth, A.~Maquet, R.~Ta{\"\i}eb, A.~L`Huillier, and A.~S. Kheifets,
\newblock Physical Review A {\bf 85}, 053424 (2012).

\bibitem{schoun2014}
S.~B. Schoun, R.~Chirla, J.~Wheeler, C.~Roedig, P.~Agostini, L.~F. DiMauro, K.~J. Schafer, and M.~Gaarde,
\newblock Physical Review Letters {\bf 112}, 153001 (2014).

\bibitem{caillat2011} 
J.~Caillat, A.~Maquet, S.~Haessler, B.~Fabre, T.~Ruchon, P.~Sali\`{e}res, Y.~Mairesse, and R.~Ta\"{i}eb,
\newblock Physical Review Letters {\bf 106}, 093002 (2011).

\bibitem{neppl2012attosecond}
S.~Neppl, R.~Ernstorfer, E.~M. Bothschafter, A.~L. Cavalieri, D.~Menzel, J.~V.
  Barth, F.~Krausz, R.~Kienberger, and P.~Feulner,
\newblock Physical Review Letters {\bf 109}, 87401 (2012).

\bibitem{cavalieri2007attosecond}
A.~L. Cavalieri, N.~Müller, T.~Uphues, V.~S. Yakovlev, A.~Baltuska, B.~Horvath,
  B.~Schmidt, L.~Blümel, R.~Holzwarth, S.~Hendel, M.~Drescher, U.~Kleineberg,
  P.~M. Echenique, R.~Kienberger, F.~Krausz, and U.~Heinzmann,
\newblock Nature {\bf 449}, 1029 (2007).

\bibitem{zhang2009metal}
C.~-H Zhang and U.~Thumm,
\newblock Physical Review Letters {\bf 102}, 123601 (2009).

\bibitem{dixit2013} 
G.~Dixit, H.~S. Chakraborty, and M.~E. Madjet, 
\newblock Physical Review Letters {\bf 111}, 203003 (2013).

\bibitem{deshmukh2014}
P.~C. Deshmukh, A.~Mandal, S.~Saha, A.~S. Khaifets, V.~K. Dolmatov, and S.~T. Manson,
\newblock Physical Review A {\bf 89}, 053424 (2014).

\bibitem{pazourek}
R.~Pazourek, S.~Nagele, and J.~Burgd\"{o}rfer,
\newblock Faraday Discuss. {\bf 163}, 353 (2013).

\bibitem{barillot2015} 
T.~Barillot, C.~Cauchy, P~-A. Hervieux, M.~Gisselbrecht, S.~E. Canton, P.~Johnsson, J.~Laksman,
E.~P. Mansson, J.~M. Dahlstr\"{o}m, M.~Magrakvelidze6, G.~Dixit, M.~E. Madjet, H.~S. Chakraborty, 
E.~Suraud, P.~M. Dinh, P.~Wopperer, K.~Hansen, V.~Loriot, C.~Bordas, S.~Sorensen and F.~L\'{e}pine,
\newblock Physical Review A {\bf 91}, 033413 (2015).

\bibitem{madjet2008} 
M.~E. Madjet, H.~S. Chakraborty, J.~M. Rost, and S.~T. Manson, 
\newblock Journal of Physics B {\bf 41}, 105101 (2008).

\bibitem{van1994exchange}
R.~Van~Leeuwen and E.~J. Baerends,
\newblock Physical Review A {\bf 49}, 2421 (1994).

\bibitem{ciappina2007-2008}
M.~F. Ciappina, A.~Becker, and A.~Jaro\'{n}-Becker,
\newblock Physical Review A {\bf 76}, 063406 (2007);
Physical Review A {\bf 78}, 063405 (2008).

\bibitem{troullier1992} 
N.~Troullier and J.~L. Martins, 
\newblock Physical Review B {\bf 46}, 1754 (1992).

\bibitem{weaver1991} 
J.~H. Weaver, J.~L. Martins, T.~Komeda, Y.~Chen, T.~R. Ohno, G.~H. Kroll, and N.~Troullier,
\newblock Physical Review Letters {\bf 66}, 1741 (1991).

\bibitem{vos1997} 
M.~Vos, S.~A. Canney, I.~E. McCarthy, S.~Utteridge, M.~T. Michalewicz, and E.~Weigold,
\newblock Physical Review B {\bf 56}, 1309 (1997). 

\bibitem{ruedel2002oscExp} 
A.~R{\"u}del, R.~Hentges, U.~Becker, H.~S. Chakraborty, M.~E. Madjet, and J.~M. Rost,
\newblock Physical Review Letters {\bf 89}, 125503 (2002).

\bibitem{taylor1976}
K.~T. Taylor and P.~G. Burke,
\newblock Journal of Physics B {\bf 9}, L353 (1976).

\bibitem{korica2010}
S.~Korica, A.~Reink\"{o}ster, M.~Braune, J.~Viefhaus, D.~Rolles, B.~Langer, G.~Fronzoni, D.~Toffoli,
M.~Stener, P.~Decleva, O.~M. Al-Dossary, U.~Becker,
\newblock Surface Science {\bf 604}, 1940 (2010).

\bibitem{mccune2008}
M.~A. McCune, M.~E. Madjet, and H.~S. Chakraborty,
Journal of Physics B {\bf 41}, 201003 (2008).

\bibitem{cooper1962photoionization}
J.~W. Cooper,
\newblock Physical Review {\bf 128}, 681 (1962).

\bibitem{frank1996}
O.~Frank and J.~-M. Rost,
Zeitschrift f\"{u}r Physik D {\bf 38}, 59 (1996).

\bibitem{wigner1955smith1960}
E.~P. Wigner,
\newblock Physical Review {\bf 98}, 145 (1955); F.~T. Smith,
\newblock Physical Review {\bf 118}, 349 (1960).

\bibitem{ivanov2011accurate}
M.~Ivanov and O.~Smirnova,
\newblock Physical Review Letters {\bf 107}, 213605 (2011).

\bibitem{dahlstrom2012theory}
J.~M. Dahlstr{\"o}m, D.~Gu{\'e}not, K.~Kl{\"u}nder, M.~Gisselbrecht,
  J.~Mauritsson, A.~L`Huillier, A.~Maquet, and R.~Ta{\"\i}eb,
\newblock Chemical Physics {\bf 414}, 53  (2012).

\bibitem{magrakvelidze2014}
M.~Magrakvelidze, M.~E. Madjet, G.~Dixit, M.~Ivanov, and H.~S. Chakraborty, {\em submitted}.

\bibitem{fano1961}
U. Fano,
\newblock Physical Review {\bf 124}, 1866 (1961).

\bibitem{landau}
L.~D. Landau, and E.~M. Lifshitz, Quantum Mechanics: Nonrelativistic Theory 3rd ed., Pergamon Press Oxford, 1977 (p. 136).

\bibitem{gunnarsson1976}
O.~Gunnarsson and B.~Lundqvist,
\newblock Physical Review B {\bf 13}, 4274 (1976).

\bibitem{chakraborty}
H.~S. Chakraborty, R.~G. Nazmitdinov, M.~E. Madjet, and J.~-M. Rost,
arXiv:cond-mat/0111383.  

\bibitem{jaenkaelae2011}
K.~J\"{a}nk\"{a}l\"{a}, M.~Tchaplyguine, M.~-H. Mikkel\"{a}, O.~Bj\"{o}meholm, and H.~Huttula,
\newblock Physical Review Letters {\bf 107}, 183401 (2011).

\end{thebibliography}

\end{document}